# Should ChatGPT Write Your Breakup Text? Exploring User Needs and Perception of AI's Role in Ending Relationships

YUE FU, Information School, University of Washington, US
YIXIN CHEN, Information School, University of Washington, US
ZELIA GOMES DA COSTA LAI, Human-Centered Design and Engineering, University of Washington, US
ALEXIS HINIKER, Information School, University of Washington, US

Relationships are essential to our happiness and wellbeing, yet their dissolution—the final stage of a relationship's lifecycle—is among the most stressful events individuals can experience, often leading to profound and lasting impacts. With the breakup process increasingly facilitated by technology, such as computer-mediated communication, and the likely future influence of generative AI (GenAI) tools, we conducted a semi-structured interview study with 21 participants. We aim to understand: 1) the current role of technology in the breakup process, 2) the needs and support individuals seek during this time, and 3) how GenAI might address or undermine these needs. Our findings show that people have distinct needs at various stages of breakups. While currently technology plays an important role, it falls short in supporting users' unmet needs. Participants envision that GenAI could: 1) aid in prompting self-reflection, providing neutral second opinions, and assisting with planning leading up to a breakup; 2) serve as a communication mediator, supporting wording and tone to facilitate emotional expression during breakup conversations; and 3) support personal growth and offer companionship after a breakup. However, our findings also reveal participants' various concerns about involving GenAI in this process. Based on our results, we discuss the potential opportunities, design considerations, and harms of GenAI tools in facilitating people's relationship dissolution.

Additional Key Words and Phrases: Generative AI, AI-Mediated Relationship, breakup, interview,

## 1 INTRODUCTION

Close relationships are essential to people's well-being [12]. Relationship dissolution, the final phase of a relationship's life cycle, can have profound and long-lasting effects on the people in the relationship and their future connections with others [49]. As one of the most stressful events in an individual's life [26], relationship dissolution is often associated with negative emotions [61], reduced self-esteem [27, 67], and reduced psychological well-being [49]. On the positive side, strategically navigating this stage can alleviate the stress associated with the dissolving relationship [75], prepare people for future relationships [87], and lead to personal growth [87].

Technology increasingly plays a mediating role in the process of dissolving a relationship. The advent of Computer-Mediated Communication (CMC), such as messaging apps and social media, has transformed people's direct and indirect interpersonal communication [69, 99]. The affordances of these tools, such as Facebook's "relationship status," the ability to "unfriend," and the ability to restrict profile access [70], significantly influence how users experience changes in their close relationship [54, 72]. A 2015 Pew Research report documented that 27% of teenagers had broken up with someone via text, 29% via phone call, and 16% via social media features (e.g., changing relationship status, messaging, posting updates) [55]. Moreover, a significant number of individuals rely on technology for advice on ending relationships. The collaborative information-sharing platform "wikiHow" [3] hosts numerous advice articles on breakups and recovery

Authors' addresses: Yue Fu, chrisfu@uw.edu, Information School, University of Washington, Seattle, Washington, US; Yixin Chen, Information School, University of Washington, Seattle, Washington, US, ; Zelia Gomes da Costa Lai, Human-Centered Design and Engineering, University of Washington, Seattle, Washington, US, ; Alexis Hiniker, Information School, University of Washington, Seattle, Washington, US, .

that have millions of views [2]. Although some people still prefer to end relationships via in-person interaction, many have moved to online channels.

With the rise of generative AI (GenAI) tools powered by large language models (LLMs), people's use of technology is expanding beyond tasks of searching for and reading advice columns on the Internet or sending messages. Recent GenAI applications extend the traditional role of technology and expand to influencing interpersonal relationships. For instance, newly developed GenAI tools claim to be able to serve coaches [88], psychotherapists [5], and even intimate companions [59]. Supportive AI tools that can transform users' language to be more acceptable and engaging are actively being studied for incorporation into dating apps [39, 76]. Given the various functions GenAI offers, it is natural to ponder how AI might influence the final stage of relationships—the dissolution process.

Thus, our work seeks to understand how people currently use digital technology and how they envision emerging GenAI tools during a breakup. The way people incorporate these technologies can significantly affect their relationship dissolution process and, subsequently, their well-being, self-perception, and future relationships. We also aim to understand the opportunities, concerns, harms, and ethical considerations people face when using technologies in this sensitive domain, particularly regarding the role of emerging AI technology. Specifically, we ask:

- **RQ1:** What role, if any, does technology currently play in relationship dissolution?
- **RQ2:** What specific needs do people have throughout the relation-dissolution process?
- **RQ3:** How, if at all, might technology, including GenAI tools, support or undermine these needs?

To answer these questions, we conducted 21 semi-structured interviews with people who have ended a close relationship in the past three years (12 people ended a romantic relationship and 9 people ended a close friendship). We explored participants' thoughts, feelings, and needs leading up to, during, and after breakups. We also asked about the role technology currently plays during breakups, participants' vision for the role that GenAI tools might play, how these tools could better support their needs, and participants' concerns about what might go wrong.

Participants described many specific needs throughout the breakup process. Specifically, leading up to a breakup, participants needed support for self-reflection, supportive and honest external perspectives, and—if they decided to move forward with the breakup—encouragement to do so. During the breakup, they needed to navigate heated conversations and maintain direct and clear communication. After the breakup, the focus shifted to emotional validation, seeking companionship, severing digital connections, and managing feelings of embarrassment. Although they explained that technology currently plays an important role in the breakup process—used for gathering information, planning actions, handling communication, and blocking contact—they further explained that it does not meet the complex needs individuals face during breakups. Participants envision GenAI tools could fill the gap by aiding understanding and reflection, providing safe and anonymous support, serving as a communication mediator, supporting companionship and personal growth, and facilitating recovery. Based on our findings, we discuss the potential opportunities, harms and design consideration in supporting people's relationship dissolution in the age of GenAI.

To summarize, our contributions are:

- We surface individuals' needs during the distinct stages of relationship dissolution, highlighting a predictable timeline of relationship dissolution.
- We examine participants' perceptions of technology, particularly GenAI, and its potential to support or hinder individual needs during breakups.
- We discuss the opportunities, harms, and ethical considerations of using the emerging GenAI in the sensitive process of relationship dissolution.

## 2 RELATED WORK

### 2.1 Relationship Dissolution and Its Impact

Interpersonal relationships, particularly close ones, are central to our emotional and psychological well-being [42]. They provide essential support [32], foster personal growth [52], and significantly contribute to life satisfaction [33]. Consequently, relationship dissolution, the process of ending a relationship voluntarily by at least one partner [8], can have profound and multifaceted effects on individuals [43]. Unlike a series of discrete events, relationship dissolution is typically a gradual process characterized by various stages and emotional transitions. [23, 46, 53].

Extensive research in psychology and social science has explored the impact of breakups, revealing a wide range of negative effects. Relationship dissolution can lower self-esteem [67] and increase the risk of depression [62]. Individuals experiencing a breakup often report acute psychological distress, including significant emotional swings, outbursts of irritation and anger, and heightened startle responses triggered by memories or reminders of their ex-partner's behavior [19]. Additionally, some individuals exhibit avoidance behaviors, feeling numb and disinterested in the world around them [19]. The breakup process is especially hard for those who relied on their former partner as a key component of their social support network [31]. Losing a partner also leads to shifts in self-concept, as individuals struggle to redefine themselves and understand their identities [84].

Despite the predominantly negative impacts, research indicates that breakups can also lead to positive outcomes, such as personal growth and increased life satisfaction [41, 79]. Tashiro and Frazier found that individuals undergoing a breakup could identify an average of five positive changes that they could make to improve their romantic lives and future relationships after a breakup, including personal growth, better partner selection, improved relationship skills, greater relationship expectations, and enhanced appreciation for social support [87]. Additionally, ending an unhappy relationship can relieve stress [7, 75], allowing individuals to pursue more fulfilling and less stressful connections.

In this study, we focus on the dissolution of close relationships. By examining the experiences and needs of individuals during different stages of a breakup, we aim to uncover how technology can support or hinder relationship dissolution.

### 2.2 Digital Technology in Relationship Dissolution

The CHI and CSCW communities have extensively researched the role of technology in interpersonal relationships. Previous studies have primarily focused on how socio-technical systems can build, support, and maintain relationships [28, 94, 98], design interactive technology to foster meaningful communication [44], support long-distance relationships [64], and address privacy and security issues in communication technologies [38].

However, the widespread use of digital technology aimed at fostering connection can complicate how people navigate relationship dissolution [61, 69]. Post-breakup, individuals often experience strong negative emotions associated with their use of digital technologies. Through platforms like social media, people express distress, anger, and vengefulness, while also crafting highly curated profiles that suggest they are fine without their ex-partners, even when this is not the case [69]. This can lead to feelings of regret over their online posts [69]. Additionally, some individuals engage in stalking ex-partner's social media profiles, especially those hoping for reconciliation, which hinders their recovery [85]. Recommendation algorithms can exacerbate emotional stress by promoting ex-partner related content without considering the current relational context of a breakup [72]. Managing digital possessions post-breakup also presents challenges. Digital memories spread across multiple devices, applications, web services, and platforms can complicate the sensitive period following a breakup as individuals struggle to dispose of them [80]. Most people either keep or discard everything impulsively, lacking the ability to dispassionately evaluate their digital possessions [80].

Shared digital possessions stored on an ex-partner's devices are particularly difficult to manage, as previously shared sensitive content—such as sexual content—can become sources of discomfort, anxiety, or fear due to the risk of being leaked or misused by a no longer trusted ex-partner [20].

Moreover, individuals need to reconstruct their identities after a breakup [71]. Existing technology tools fall short in supporting competing desires to present authentic past and future online identities [70]. Mutual connections on social media, including friends and ex's families, often create barriers to completely removing an ex from one's online social network. In some cases, individuals choose to quit or deactivate their social media accounts altogether to sever all connections with ex-partners [72].

We build on the literature by exploring people's experiences and needs during different stages of a breakup and their perception of the current use of technology in these stages.

### 2.3 GenAI for Interpersonal Relationship

GenAI is increasingly being integrated into various applications to improve communication and enhance human relationships. These technologies promise increased accessibility and personalized support for areas where demand often exceeds the availability of traditional resources. For instance, Replika, a virtual AI companion, aims to support people by being available whenever and for whatever purposes they might need it [1]. With the rapid advancement of GenAI technologies and growing societal interest, these applications have become increasingly popular in recent years. Replika reported two million total users, 250,000 of whom were paying subscribers in 2023 [30]. A Chinese chatbot named Xiaoice claims to have hundreds of millions of users and a valuation of about $2 billion, according to a recent funding round [100]. And numerous GenAI-powered dating apps offer message suggestions for those hesitant to start or respond to conversations [76, 97].

Today, these GenAI tools are increasingly used to mitigate the embarrassment, friction, and ambivalence that can arise in close relationships. Applications are designed to assist with initiating conversations [39], generating flirtatious responses [10], and simulating conversations for practicing communication with potential partners [17]. Beyond communication support, social AI chatbots can provide companionship to help alleviate loneliness [86], offer empathetic and validating responses [57], and deliver mental health interventions that can reduce depression symptoms [35].

Despite these potential benefits, the adoption of AI in personal communication and relationships introduces several challenges and potential harms. Ethical considerations and practical limitations have emerged as significant concerns. Research indicates that users perceive AI-generated text as less trustworthy [40] and believe that communication partners using AI appear less cooperative and affiliative [37]. Furthermore, AI-powered communication tools often alter users' communication styles, produce generic responses, and fail to accurately convey users' intended messages [29, 36].

Research has also shown the inability of current chatbot-based AI companions to recognize and appropriately respond to signs of distress, which limits their effectiveness in providing meaningful emotional support [22]. In addition, some users of AI companions may form maladaptive bonding and overdependence on virtual companions, potentially undermining users' ability to form healthy human relationships [51]. Moreover, unmonitored GenAI chatbots may pose severe consequences, such as facilitating suicidal ideas. Several cases have linked suicides to interactions with GenAI-powered chatbots [78, 96], particularly among younger populations who are still developing their understanding of human relationships and constitute the primary user base for AI companions like Replika [78]. These AI chatbots often lack appropriate safeguards to protect user safety [13, 21].

Critics argue that GenAI applications can exacerbate problems seen in existing social media by personalizing information in ways that reinforce user biases, undermine listening skills, and extend user engagement on these platforms [91]. By integrating AI into intimate conversations [56], GenAI collects more user data than any previous social media platform, raising concerns about data privacy and the commercialization of personal information.

Building on the literature, our study examines relationship dissolution—a period of significant emotional and psychological distress—to understand the needs individuals have during this sensitive time and to investigate whether emerging GenAI technologies can support or undermine users in managing relationship dissolution. By exploring how users wish GenAI technologies to assist them in this process, our research contributes to the ongoing discourse on the role of AI in facilitating or impeding communication, relationships, and mental health.

## 3 METHOD

We conducted semi-structured interviews with 21 participants to understand their perspectives on technology's role in breakups. We included both participants who had ended romantic relationships and participants who had ended close friendships, because the strength of an individual's closest relationships—whether romantic or platonic—is the strongest predictor of that person's well-being [60]. In both cases, these close relationships offer protective effects against adverse life events and enhance health and well-being by providing emotional support, companionship, and greater life satisfaction [12]. By studying breakups in both contexts (rather than in romantic contexts alone), we aimed to gain a more comprehensive understanding of the dissolution of these critically important close relationships.

Previous research shows there are some key differences between romantic and platonic breakups: romantic breakups are more likely to involve intense distress due to deeper interdependence and intimacy, sometimes leading to depression [11, 93]. And friendship breakups may end implicitly without direct discussions and are often perceived as mutual decisions, especially among adolescents [95]. Despite these differences, research also shows that both types of breakups elicit very similar emotional experiences [14, 68], behavioral patterns, and coping strategies (including common communication strategies and patterns of attributing blame) [6, 9, 25, 47, 48].

Recognizing that the term "breakup" often connotes romantic relationships—and indeed some studies use "breakup" synonymously with romantic breakups [48]—we realized that using this term alone in our recruitment materials might fail to attract individuals who had ended close friendships, leading to underrepresentation. To address this and capture a broader perspective on how people navigate the dissolution of close relationships, we created separate recruitment materials using the specific terms "ending a friendship" and "ending a romantic relationship." This recruiting approach ensured we reached participants from both groups, facilitating an inclusive exploration of close-relationship breakup experiences. In total, we recruited 12 who had undergone romantic relationship breakups and 9 who had experienced close friendship breakups. We piloted the study with two participants from the academic institution of the authors and included this pilot data in the paper (pilotA and pilotB).

### 3.1 Participants

We recruited 21 participants through professional and academic Slack channels and physical posters at the authors' institution. We advertised as a study of relationship breakups. We created two posters, one for recruiting romantic relationship breakups and another for recruiting close friendship breakups. The initial screening survey asked about the timing of the breakup, whether participants initiated it, the communication method used for the breakup (e.g., in person, phone call, text messages), the reasons behind the breakup, and the duration of the breakup process. Additionally, for the close friendship group, participants were asked to evaluate the significance of the friendship, ranging from "most important" to "not important." To qualify, participants were required to have had a breakup within the last three years; those in the close friendship group had to rate the friendship as at least moderately important. We aimed to recruit participants with a variety of responses to these screening questions and thus reflecting a diverse set of breakup experiences. Recruitment and interviews took place from October 2023 through early December 2023. Our analysis showed data saturation after 21 participants. We present the demographic data in Appendix A. Participants were compensated $20-$30 based on interview length ($20 for an 30 mins interview, and $30 for an 45 mins).

### 3.2 Materials

Initially, one researcher drafted 12 potential interview questions, which were then refined and expanded upon by the research team. The protocol was tested with pilot participants (pilotA from romantic relationship and pilotB from close friendship). Piloting showed similarities in the breakup processes of romantic and close relationships, typically involving three stages demarcated by "relationship talk": the phase leading up to the breakup, the breakup, and the phase after the breakup. We define the term "relationship talk" as instances when people communicate their intention to end their relationship with either romantic partners or close friends. Based on these findings, we revised our interview protocol. The protocol for both romantic and close friendship groups remains the same, with only minor alterations in relationship terms. The final protocol includes four parts. The first part asked descriptive questions such the duration and closeness of the relationship and the reasons for breakups. The second we explored participants experience and communication during the three stages of breakup. During the third part, we asked participants to reflect on their breakup process, probing faced challenges, support needed, and their overall well-being. For the last part, we invited participants to envision technologies that could support breakups. We probed their perception and ideas of using an AI to support their experience, though we did not specifically detailing what the AI is. Participants were encouraged to freely speculate. At last, we presented participants several general design vignette, such as an AI analyzing text messages, and an AI assisting in crafting empathetic communications.

### 3.3 Procedure

We emailed qualified participants the study information and offered them two options: 1) a one-on-one interview with interviewer, conducted either online or at the authors' institution, or 2) a one-on-two interview, accompanied by a friend with similar experiences. We offered the one-on-two interview option because previous research suggests that dyadic interviews can yield data not accessible through individual interviews, as participant pairs can prompt and cue each other's memories and reflections [63]. Joint interviews provide a supportive environment where participants feel more comfortable sharing personal experiences and can expose differences and similarities in their experiences [81]. We predicted that some participants might prefer discussing breakups in this format, as friends who know each other well can provide support and can encourage each other to share both similar and differing perspectives, potentially generating richer data compared to one-on-one interviews with a researcher alone.

Three participants opted for the one-on-two interview format with a friend who had also experienced a breakup previously. After each 1-on-2 interview, the friends completed a screening survey. For the one-on-one interviews, we followed our semi-structured interview protocol. In the one-on-two interviews, we mostly adhered to the same protocol, asking each member of the dyad the same questions. However, the process was more flexible in these sessions. The pair usually prompted each other fluidly, sharing experiences, rapport, and reflections, and we followed their lead to facilitate the discussions. The interviews were designed to last 30 minutes for one-on-one sessions and 45 minutes for one-on-two sessions. All interviews were audio-recorded and transcribed. We anonymized and securely stored all audio recordings. Our institutional review board (IRB) reviewed and granted an exemption for this study.

## 3.4 Data Analysis

We used Reflexive Thematic Analysis [15, 16] to analyze our data, as it allows for a flexible and in-depth exploration of participants' experiences while acknowledging the researcher's active role in interpreting the data. We began analyzing the data concurrently while interviewing. The research team read the transcripts and revisited the audio recordings. Three team members met weekly to discuss their initial understandings and findings from the interviews.

Initially, we divided the dataset based on the two relationship types—romantic relationship breakups and close friendship breakups. However, after the initial round of reading and group discussions, we discovered that participants' experiences were largely similar across the relationship types, aligning with the literature [50, 92]. Participants shared comparable reasons, strategies, emotional reactions, and needs during various stages of relationship dissolution. Thus, for subsequent rounds of coding and analysis, we combined the data from both relationship types and coded them together.

After this initial data familiarization phase, one team member systematically generated initial codes using the qualitative analysis tool Delve[1]. The team collaboratively discussed these codes, combining, revising, and refining them. Guided by the research questions, two authors then independently condensed and organized the data and codes into initial themes such as reasons and coping strategies for the breakups, participants' needs across different stages of the breakup timeline, their feelings and thoughts during the process, communication methods, current technology use during the breakup, their envisioning of GenAI technology's role in this process, and their concerns of using technology. We met weekly to compare our individual themes, discuss discrepancies, and reach consensus on the thematic concepts.

In the next round of analysis, one team member revisited and reviewed the coded data, merging related themes and discarding those not relevant to the research questions. The research team met frequently to discuss and build consensus on the themes. During these discussions, we agreed that organizing and presenting the data by breakup stages was most appropriate. This decision was informed by existing literature illustrating various stages for relationship dissolution [53] and emerged from our data, which showed that participants have different needs, technology uses, and envision different technological support at different stages of breakups. After refining the themes, one team member applied the updated thematic structure to the entire dataset of interview transcripts.

Subsequently, one author extracted the themes, codes, and related interpretations with participants' quotes from Delve and systematically documented and organized the data into a Word document. The lead author then wrote a first draft of the Results section of this paper based on this organized material, which the research team edited collaboratively.

## 4 RESULTS

In our study, participants shared various needs for a breakup, identified different roles of current technology, and envisioned opportunities for AI technology support. Participants 1 to 12 experienced a romantic relationship breakup, and participants 13 to 21 experienced a platonic close friendship relationship breakup. We employed a labeling system to denote the two types of close relationships. Close friendship participants received a "-F" suffix added to the participant ID (e.g. "P15-F"), whereas for romantic relationship participants, we used participant ID without a suffix. The labeling system can help readers understand participants' quotes in the context of relationship types. We summarized participants' identified needs and envisioned GenAI's potential support, and their concerns during each stage of the breakup process in Fig. 1.

---

[1] Delve: https://delvetool.com/

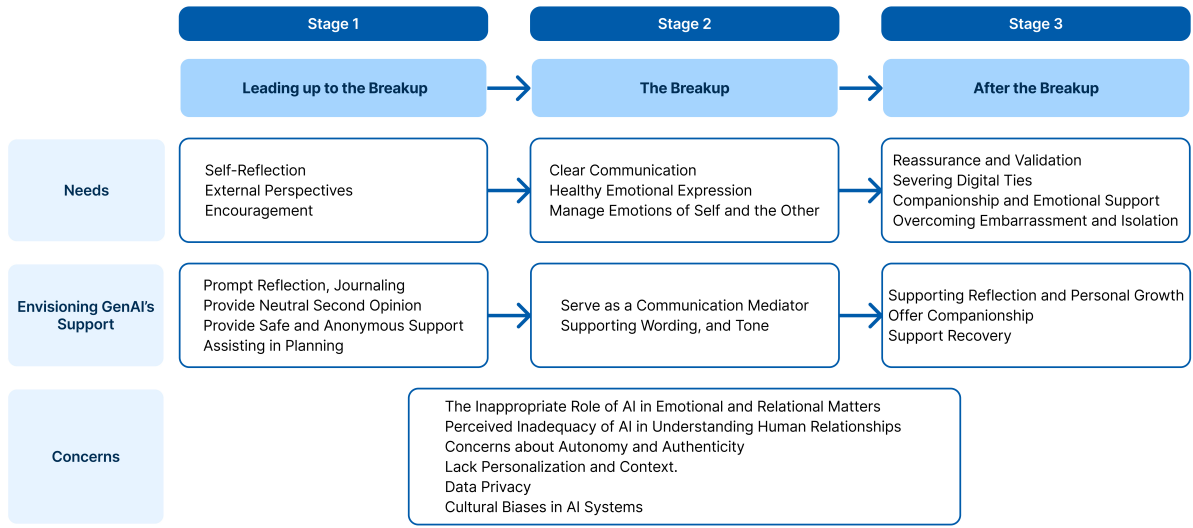

Fig. 1. Summary of users' needs, potential GenAI design space in supporting relationship dissolution, and users' concerns using GenAI across three stages of the breakup process.

## 4.1 Stage 1: Leading up to the Breakup

Participants described a period of introspection and re-evaluation of a relationship that occurs before a breakup. During this stage, they sought a better understanding of their relationship, expressing a need for clarity, self-assurance, and honest feedback. To achieve this, participants engaged in internal reflection and sought external perspectives from friends and family, with these processes frequently mediated by technology.

*4.1.1 Participants' Experiences and Needs in Stage 1.* Participants engaged in internal self-reflection to assess their feelings and the state of their relationship, often experiencing confusion and uncertainty. One participant articulated the need for internal clarity, saying, "*I'm seeking some kind of support that, I'm not doing anything wrong...At that time, I was just quite confused about myself and everything*" (P14-F). Similarly, P3 struggles to trust their own judgment with their partner nearby, "*my ex was extremely verbally affectionate...it was really hard for me to trust my own perspective enough and also trust my feelings as being a valid judgment*" (P3). Some participants considered structured self-reflection methods like journaling to help process their emotions. One participant said, "*[Journaling] might have been able to help me in a sense of sorting out my thoughts in a more organized way and reflecting on the different emotions I had and why I felt them*" (P15-F).

In addition to internal reflection, participants also sought external perspectives to gain insights they felt unable to see themselves. P17-F described seeking advice from friends, saying, "*I was telling them about how I wasn't feeling heard, I wasn't feeling respected...they were like, 'yeah, I don't think you should be friends with her anymore.'*" Another participant stressed the importance of outside viewpoint, saying, "*[my friend] had really cautioned me...she saw things that I didn't*" (P15-F). Participants valued honest feedback from trusted individuals, recognizing that being within the relationship made it difficult to judge fairly, as P12 explained, "*I would think that their [friends and family members'] advice is really helpful because that helps me. They are standing from a third-person perspective, and they can really see how our relationship goes. During that time I was part of the relationship, and it's really hard to make a really fair judgment*" (P12). However, participants also explained that honest and compassionate external feedback was not always available. Some loved ones withheld their true feelings until after the breakup. Participants said things like, "*nobody liked him, but also nobody would tell me that until after I broke up with him*" (P11). Others explained that their friends sometimes lacked sufficient context or experience to empathize, leaving the participant with the feeling that "*they won't completely understand*" (P5).

Participants also faced nervousness and uncertainty about initiating the breakup conversation. They struggled with how to express their decision and feared potential repercussions. P11 admitted, "*I didn't know how to tell this person that I didn't want to be with them anymore…dragging this [the process of breaking up] on forever and ever*" (P11). Another questioned, "*What's the right thing to say? How do I get my message across?*" (P10). Similarly, P15-F echoed these concerns: "*What do I say to her? When and how do I say it?*" (P15-F). Participants explained that, at times, encouragement from friends and family was important to giving them the courage to initiate the breakup. As one shared, "*I had multiple conversations with my friends…getting up the nerve to break up with this person*" (P2). Others found accountability helpful, "*I told my mom that I wanted to break up with him…then it was this accountability thing,*" and "*Telling them [two close friends and a sister] held me accountable [since] they were expecting [me to initiate the breakup].*" (P1). In these cases, the support system provided the encouragement and sense of responsibility needed to move forward.

To summarize, participants' needs in Stage 1 include:

- **The need for self-reflection and clarity about the relationship:** Participants expressed a need to assess their current relationship status, focusing on understanding their own feelings and the future prospects for the relationship.
- **The need for honest external perspectives:** Participants emphasized seeking unbiased opinions from trusted friends and family to gain insights they might have missed due to their emotional involvement.
- **The need for support and encouragement to initiate the breakup if they choose to proceed to Stage 2:** Participants said they value having support to overcome nervousness and uncertainty, seeking guidance on how to communicate their decision effectively and the confidence to proceed.

*4.1.2 Technology's Current Role in Stage 1.*
**A Source of Advice, Information and Community Support.** During the stage leading up to a breakup, participants often turned to technology for advice, information, and community support. They used online resources to seek guidance, validate their feelings, and learn about others' experiences in similar situations. One participant mentioned searching the internet for relationship advice, saying, "*It could give me something that is contextual to me. So sometimes I would Google about these things, like breakups and how it would go and how you do it rightly*" (P5). Others found comfort and a sense of solidarity by reading about others' breakup experiences online. P8 explained, "*It is useful using the internet and reading other people's experiences…because it helped show me that I wasn't the only one who has been through a breakup.*" Similarly, P11 turned to TikTok to learn from others' breakup journeys, "*Before I did the breakup, I was looking a lot at TikTok and what other people had done, how other people had broken up with their partners, or seeing other people go through their breakups was really nice because I knew that I wasn't alone…Seeing other people's stories and experiences really helps*" (P11). Reading other people's stories allowed participants to feel less isolated and more understood, providing emotional support during a confusing time.

Some participants actively sought advice and perspectives from online forums. P3 described posting Reddit, saying, "*I wrote, like, a huge paragraph and I posted it on Reddit…The only advice they [users on Reddit] gave were like, you should* *probably break up with him*" (P3). As this example illustrates, participants at times used online platforms to crowdsource opinions and gain external perspectives that they might not receive from their immediate social circles.

Engaging with online resources sometimes helped participants realize truths they already sensed internally. Technology served as a tool for self-reflection and affirmation, helping participants acknowledge their true feelings. For example, one participant described turning to Google, saying, "*The more I was looking into those resources, the more I was imagining that life without him, the more I realized, oh, I'm looking into this because this is the answer I already know*" (P9).

**Technology as a Catalyst for Surfacing Relationship Issues.** In some instances, technology acted as a catalyst by revealing critical information about the relationship, prompting participants to reconsider their partnerships. One participant shared that accessing the partner's phone led to a pivotal moment:

> "*The only reason I found out how she felt was because we were at a party. We were both drinking, and then I went on her phone, and I was like, okay, I'm going to try and figure out…And maybe that's a little sneaky. I shouldn't have done that, but I did it. After I read through her phone and stuff, I jumped to the conclusion that she wanted to break up, but she didn't know how to do it*" (P10).

Others described stumbling onto online content accidentally that made them question the relationship. For example, P13-F discovered a friend's misconduct through social media posts of others. P13-F mentioned, "*I woke up to [find] eight or nine posts and a bunch of people tagged me in the comments. As soon as I woke up to all that content, I saw the words and I saw the screenshots and the evidence [on social media]. In that moment, I was like, I'm not going to assess [the relationship] here with him*" (P13-F). This unexpected revelation on social media prompted her to evaluate the friendship and move toward ending it.

These accounts show that technology can play an active role in the breakup process—not only as a tool for deliberately seeking advice but also as a means through which critical information is unexpectedly uncovered. Whether through intentional searches or accidental discoveries, pivotal technology-facilitated realizations influenced participants' decisions to end their relationships.

*4.1.3 Envisioning GenAI's Support for Stage 1.* Participants envisioned several ways in which technology, particularly AI, could better support them in Stage 1 by helping them understand themselves, their relationships, and the potential need to initiate a breakup.

**Supporting Self-Reflection.** Participants felt that AI could assist in understanding their personal feelings and relationship dynamics. P15-F believed that AI could "*definitely be helpful in understanding your personal feelings*" (P15-F). P12 mentioned AI could prompt introspection without giving direct commands, "*AI could ask me questions that help me really realize that this guy is not the right for me, instead of just telling me directly that you should do this*" (P12). Participants suggested a feature where AI could provide relationship feedback and interpretation on messages, saying something like, "*I will put [into ChatGPT] 10 things went like this today, what should I do then?*" (P7), and another participant wanted help interpreting what the other person was saying, "*[help] interpret what [the breakup party] was saying to me*" (P6). These reflections indicate that participants saw AI as a supportive tool for gaining clarity about the dynamics of their relationships, facilitating deeper understanding, and aiding in decision-making.

**Providing a Neutral Second Opinion.** Participants suggested turning to AI for a second opinion, free from the potential biases of friends and family. P18-F saw AI as a beneficial tool to escape the trap of overthinking alone: "*Alone *with my thoughts is a dangerous place to be. It would help have at least a second opinion of sorts*" (P18-F). Both P1 and P17-F envisioned AI helping to identify subtle issues in their relationships, such as "*signs or red flags that are very common or not obvious at the time*" (P1). Participants envisioned AI providing feedback that would not be influenced by personal relationships or emotions. One participant explained: "*If you go to your friends about it, they often only get one side of the story...[but] AI would give you a straightforward [perspective]...it would help process feelings better and decide what you want to do without acting emotionally*" (P19-F).

**Providing a Safe Place to Share Anonymously.** Participants anticipated the anonymity provided by AI being valuable, because it could enable a safe space for sharing sensitive information without the fear of judgment or betrayal. P3 shared their hesitation to discuss relationship issues openly with others: "*I would withhold a lot of personal information when talking with friends, the fights we were having, the things he had done to hurt me...I thought speaking behind his back was going against him*" (P3). Participants envisioned the anonymity provided by AI allowing them to express themselves more freely. P16-F mentioned the appeal of AI for those hesitant to seek in-person help: "*It could be a lot of help to a lot of people, especially people who don't want to take the step to go forward and talk to somebody in person. I think that factor of anonymity would be really helpful*" (P16-F). Similarly, P9 emphasized the ability to "*essentially anonymously get advice, get validation*" (P9). These examples reflect participants' perspective that AI could provide a confidential environment to explore their feelings without social repercussions.

**Assisting in Breakup Planning.** Participants recognized the potential for AI to assist in planning and structuring the breakup process. P18-F reflected on the challenges they had faced because they of a *lack* of structure: "*The process probably would've helped to have it more structured because that process just probably made a bigger mess*" (P18-F). Another participant pointed out the potential of having AI provide personalized strategies: "*I think that a lot of people struggle with finding the right process...Seeing different strategies written out algorithmically specific for your situation would be interesting*" (P2). Participants envision AI as a tool that could provide customized advice and step-by-step support to make the process smoother and less overwhelming.

### 4.2 Stage 2: The Breakup

*4.2.1 Participants' Experiences and Needs in Stage 2.* During this stage, participants often initiated a relationship talk with the other party. They described struggling with miscommunications and escalating emotions. This stage is often charged with intense emotional stress, uncertainty, and heightened tension as both parties navigate the difficult conversation of ending the relationship.

**Knowing how to Communicate Clearly.** Participants expressed a need for better communication to prevent misunderstandings and emotional escalation. As one participant described, "*I was kind of harsh in the way I said it, maybe it shouldn't have been said...she took offense, and then she became angry...eventually she said, I don't want to [talk], I need space. Don't talk to me*" (P15-F). Similarly, P2 described initiating a breakup and being met with an intensely emotional reaction, saying, "*I said it within the first sentence or so, I want to break up...they [the breakup party] started really aggressively crying*" (P2). These accounts illustrate the high stakes of communication during a breakup and the likelihood of provoking intense emotional responses, which often left participants who initiated the breakup feeling uneasy and unsure of how to proceed. Participants wanted to understand how to communicate effectively without escalating the situation.

Participants consistently noted the value of being direct and clear in their breakup communication. Vague or indirect communication led to exhausting conversations that circled without resolution, causing unnecessary emotional burden and confusion, with participants saying things like, "*The first night we were up until three in the morning just kind of talking, kept bringing up the things aren't going well…Back and forth for hours was really exhausting when you can say upfront and kind of get it out of the way*" (P1). The same participant continued, "*I feel like being more direct would've been good*" (P1). A number of participants echoed this sentiment, saying things like, "*she wasn't very specific. I ended up reiterating things I had in the past…But she was very frustrated*" (P15-F). Participants emphasized the importance of being "*open and honest to people with how I'm feeling*" (P17-F) and said that they "*wanted to be clear with my emotions*" (P2).

**Emotional Attunement.** Another recurring theme was the need for emotional self-control and awareness of the other person's emotions during breakup conversations. Without emotional awareness, control, and healthy expression, conversations could easily break down and cause emotional trauma for both parties. One participant described, "*She was really emotional, we're both really emotional…We're trying to be nice, but on calls, it just kind of blurted it out*" (P17-F). Another participant expressed regret over sending numerous emotional texts: "*I sent eight or nine texts. That is something I could have done better, cut down the number of texts I sent my half, especially some of the texts where I was like, I know you're online, so pick up [the phone]…I wish I had not acted out that frustration*" (P13-F). Participants also recognized the need for guidance in appropriately expressing emotions: "*When I get angry or frustrated, my face goes blank. I take on this kind of battle within myself of what should I say right now? What is reasonable to say and how much of how I'm feeling is appropriate to communicate?*" (P15-F). This highlights participants' struggle to manage their emotions and communicate effectively during this emotionally charged stage.

Some participants mentioned a need for a facilitator who could help navigate the conversation and prevent misunderstandings that could unnecessarily escalate tensions, saying things like, "*someone needs to facilitate that to be calm, cool, collected*" (P10). P10 then elaborated, saying, "*I wish we would've been more calm…maybe we could bring in a third mutual friend to talk things out*" (P10).

To summarize, participants' needs in Stage 2 include:

- **The need for guidance about communicating clearly:** Participants wanted assistance in communicating their intentions clearly to avoid misunderstandings, confusion, and prolonged emotional distress.
- **The need for guidance on healthy emotional expression:** Participants sought assistance in expressing their emotions in a manner that is both healthy and conducive to constructive communication during a breakup.
- **The need to navigate and manage the emotions of self and other:** Participants sought support in being aware of and understanding both their own emotions and those of the other person to avoid escalating conflict or creating miscommunications.

*4.2.2 Technology's Current Role in Stage 2.* As relationships ended, previously positive aspects of technology, like the ease of staying connected, could turn negative. Technology became a significant medium for communicating the decision to end relationships, severing ties both online and offline, and erasing participants' shared digital history. We found participants used a range of technologies during their breakup process, such as Face-time (P3), texting messages (P7, 10, 13-F, 15-F, 16-F, 18-F,20-F), social media (P14-F, 17-F), email (P11). They also highlighted both the benefits and drawbacks of using technology compared to in-person interactions when ending a relationship.

**Communicating the Breakup Decision.** Technology was used to communicate participants' decision to end a relationship. Some participants described using online communication as a way to prepare for in-person breakup conversations. P1 stated, "*I've considered just using texting or messaging almost as being able to prime the breakup… So it's like you've already forced yourself to start the conversation over the text and then you're ready to have it in person*" (P1). This approach allowed them to ease into a difficult conversation by gradually introducing the topic of breakup, as one participant shared, "*I tried to hint at it through texting. I made it kind of easier to ease into the conversation. I think he understood that too*" (P16-F). By initiating the discussion online, participants could raise the subject with less immediate emotional pressure, setting the stage for a more in-depth in-person conversation.

For the actual breakup messages, participants mentioned online communication provided them with a sense of empowerment, control, and freedom to express their stories without interruptions usually experienced during in-person communication. P3 said: "*Being able to remove yourself from [in-person communication] and at least control the flow of messages, that's so much power*" (P3). Texting offered a sense of safety over face-to-face interactions, as P3 further explained, "*It's more physically threatening [to communication in person]…For me, it was really important that it was over text in the end*" (P3). The asynchronous nature of online communication allowed participants to focus on their emotions and articulate their thoughts more clearly. P3 noted, "*Whenever we would actually talk on the phone, I would never be emotionally stable or strong enough to really address the points directly…It was empowering [to communicate by texting letters], I could take my time and make sure I was comfortable with what I was going to say*" (P3). Similarly, another participant found it easier to focus on their feelings via text: "*It's easier to get tunnel-visioned on your emotions and express how you feel*" (P16-F).

Another benefit participants found in online communication was the ability to craft clearer and more thoughtful messages. Participants mentioned online communication made easier to "*iterate*" (P20-F), saying it "*allows you to read your message multiple times*" (P10). This afforded them the opportunity to refine their words and ensure their message conveyed the intended meaning. P4 explained that online communication could reduce emotional manipulation: "*Tone of voice is very important in people making decisions. This person [the breakup party] they would yell at me or snap at me, the tone of voice change. It would freak me out*" (P4). In this instance, communicating the breakup online prevented the more vocal person from dominating the conversation and reduced the impact of intimidating behaviors.

Despite these benefits, some participants felt guilt about using technology to end a relationship, as societal norms often dictate that breakups should be conducted in person. One participant reflected, "*I always felt so much guilt about doing it over text message. Everyone always says you have to at least do it on the phone. If you're a long distance, you have to do it in person. That's what I learned from listening to other people talk. That's what is expected and that's what people deserve.*" (P3). Similarly, P1 shared, "*I feel like you almost owe it to someone, or it's the right way to do things almost instead of just sending a text*" (P1). Other participants corroborated the social norm of in-person breakups, saying "*It's more genuine to break up in person*" (P2), "*Face-to-face is much better if I wanted to be more responsible*" (PilotA). These reflections indicate that while technology offers practical advantages, it can conflict with personal and societal expectations about appropriate breakup etiquette, leading to feelings of guilt or perceived irresponsibility.

**Enabling Distancing, Ghosting, and Blocking.** Participants shared how technology facilitated distancing, ghosting, and blocking. One participant mentioned using delayed texting as a strategy, "*I will kind of reply, but I will not reply within the second, which is probably the standard for him*" (P5). Similarly, PilotA spoke about procrastinating replies as a means of distancing, "*He sent me those messages and I was ignoring. And towards the end, I was kind of scared of his texts, so I just procrastinated to reply*" (PilotA). The act of blocking on social media and other platforms was frequently mentioned as a decisive step in ending relationships. Participants mentioned something like, "*The friendship ended the very next day when she unfollowed me everywhere*" (P17-F), indicating the finality of such actions. P9, blocked a partner's number and even their mother's phone after the partner attempted to contact P9 using the mother's phone. Similarly, participant P4 thoroughly blocked their partner from every communication opportunity online, saying, "*on every single app, every single communication thing, including Duolingo*". Blocking symbolizes the closure of a relationship, effectively halting all communication channels. As one participant described, "*It feels so permanent sometimes…it's like, we're stopping, we're done. Communication is over after you hit the block button.*" (P4).

However, the act of ghosting and blocking was often a very negative and emotionally distressing experience for the recipient, as described vividly by P18-F:

> "*Via text, people can ghost and block and just be done with it, then all I have are I'm just left thinking back along with my thoughts, and that's a pretty horrible feeling… It's like I was left in a room with a foul stench, [they] closed the room and left to leave the fallout zone. That felt horrible because that meant that now I don't really have any way of expressing anything. Now I'm alone in processing this, it's isolated. It's just me in my room and this social media chat. Nothing I can do except look at that message and be alone with my thoughts. And the thoughts aren't constructive because now it's just a mix of anger, sadness, a bit of self-hate*" (P18-F).

In summary, participants said that technology plays a complex role during the breakup stage. While it provided participants with tools to communicate their decisions in a controlled and emotionally safe manner, it also introduced the challenge of violating social expectations and the potential emotional harm caused by distancing behaviors like ghosting and blocking.

*4.2.3 Envisioning GenAI's Support for Stage 2.* Participants envisioned several ways in which technology, particularly AI, could support them during the breakup conversation itself. They saw AI as a potential facilitator to mediate discussions and as a tool to help them express their emotions appropriately, choose the right words, and convey the desired tone.

**Facilitating the Breakup Conversation as a Mediator.** Participants imagined AI playing a mediator role during the breakup talk, helping to maintain a constructive and calm dialogue. For instance, one participant suggested, "*A mediator would have been helpful… if someone else could have listened to our discussion, it would've been protective for me*" (P3). Similarly, another participant recognized the benefits of AI in considering both parties' perspectives and being fair, " *AI might be able to help consider both perspectives and an approach that isn't too negative or hard on either party*" (P16-F). These reflections indicate that participants wanted a neutral third party to assist in navigating the complexities of the breakup discussion, and they saw AI as a solution that could provide mediation.

**Supporting Emotional Expression, Wording, and Tone.** Many participants faced challenges in articulating their emotions and believed that AI could support them in expressing themselves more effectively during a breakup. For instance, participant P3 explained, "*I wasn't very good at typing out and expressing emotion through text.*" P10 echoed this feeling and suggested AI can help, "*I would rewrite messages… I would read it, but that doesn't sound right. I'll just keep doing that over and over again. I have some idea of how it's supposed to sound, but somebody else might think differently. So AI would be helpful for that*" (P10). P17-F mentioned the need for "*getting advice on what to say and how to phrase things.*" Others emphasized the importance of finding the right words to convey their intentions clearly and sensitively. P1 mentioned,"getting ideas of wording or specifics would help". And P2 shared similar view, "*I think at least for wording. It'd be interesting for it[AI] to take wording*" (P2). In addition, participants recognized the challenge of conveying appropriate tones during the breakup conversation. Reflecting on their past communications, some regretted being, "*blunt, not sound as compassionate as I might be feeling*" (P8) or "*a bit rude…in hindsight that was not the right* *way*" (P5). They expressed a desire to be sincere, empathetic, direct, and compassionate, predicting that AI has the potential to help. One participant shared that AI could help by "*Being direct but also balancing someone's feelings*" (P1), and another corroborated, "*It's not cold and harsh, but also still direct enough. I think [AI] could be a very interesting tool to help people*" (P2).

### 4.3 Stage 3: After the Breakup

*4.3.1 Participants' Experiences and Needs in Stage 3.* Participants experienced a range of emotions post-breakup, from sadness to relief, one participant summarized this paradoxical feeling, "*I felt happy for the experience but also contrasted with a lot of sadness for having lost someone close*" (P15-F). This emotional complexity often led to confusion and a need for reassurance about their decision and actions. Participants grappled with self-doubt, questioning whether they made the right choice. One participant expressed: "*Did I do anything wrong? I was quite confused about myself*" (P14-F), and "*Did I do the wrong thing? Was I the one who messed it up?*" (P7). To alleviate these doubts, participants often sought confirmation from friends and family, "*I reached out to a couple friends and the support was more of just validation. That was definitely what I needed and that's was helpful and just being validated*" (P20-F).

Another constant need participants mentioned was to sever digital ties with their previous partners, as lingering memories could lead to emotional stress and doubt. One participant said they want to "*make sure pictures of us do not keep coming back on my Google feed, my Facebook, and things like that*" (P13-F). The same participant also desired technology to be more intuitive to her emotional needs and stated technology should "*embed enough metadata, have enough external information that it should be able to know you what image[you] want to see*" (P13-F). This illustrates participants' desire for control over digital content to prevent unexpected reminders that could hinder their healing process.

In addition, The immediate aftermath of a breakup often led to a decline in participants' well-being, intensifying their need for companionship and emotional support. P17-F described difficulty in coping with everyday activities, while P15-F and P8 talked about emotional struggles. P13-F mentioned a negative impact on their mental and emotional state, and P5 sought professional therapy for support. Participants expressed a profound need for someone to talk to and provide comfort during this vulnerable time. P18-F gravely articulated this need, saying:

> "*I think the support I was probably looking for was just having someone to sit down with and talk about it. And maybe just a comforting touch. That would've been pretty helpful. That wasn't there, but I think I would've preferred that*" (P18-F).

Others supported this sentiment by saying, "*The kind of support that I needed was very much [someone] just being there*" (P11), and P13-F shared, "*somebody to talk to, not even for advice. The ability to say out loud to another person, the things that were in my head were helpful*" (P13-F). These reflections emphasize the critical role of emotional support and human connection in participants' recovery. However, not all participants had immediate access to such support. Due to various reasons, including geographical distance, some lacked the companionship they needed. One participant, who experienced a long-distance relationship breakup after moving to another place, shared "*I didn't have my friends around me, and I got pretty lonely pretty fast. I would say I got pretty depressed, so my social support was honestly none after I got back*" (P10). This absence of support exacerbated feelings of loneliness and hindered the healing process.

Participants also expressed a need to overcome feelings of embarrassment and shame to facilitate recovery and move forward. These emotions often hindered them from seeking help or support after the breakup. One participant explained, "*I didn't [seek help]... I don't know if embarrassed is the right word. I don't know if ashamed was the right* *word*" (P9). The difficulty in admitting they had been mistreated contributed to their hesitation: "*Ashamed for being a victim. There's a little bit of embarrassment to coming forward and admitting that you let someone treat you poorly*" (P9). P4 echoed this sentiment, stating: "*At the time [there is] this weird emotional manipulation thing, but I don't want to admit that I let myself do that…I don't want anyone to know that this is hurting me*" (P4). These accounts illustrate how stigma and self-blame can lead to isolation, making it challenging for participants to seek the support they need and move forward with their recovery.

- **The need for reassurance and validation:** Participants sought confirmation and validation from friends and family to alleviate post-breakup confusion and affirm their decision to end the relationship.
- **The need to sever digital ties:** Participants desired to cut off digital connections to their former friend or partner, including communication channels and social media feeds, to prevent unwanted reminders and reduce emotional distress.
- **The need for companionship and emotional support:** Participants expressed a strong need for companionship and someone to talk to about their emotional struggles.
- **The need to overcome embarrassment and isolation to facilitate recovery:** Feelings of embarrassment, shame, and isolation hindered participants from seeking help, revealing their need to overcome these emotions to facilitate healing and move forward.

*4.3.2 Technology's Current Role in Stage 3.* In this stage, current technology played a limited role by enabling participants to sever digital ties by unfollowing or blocking their ex-partners. Participants utilized platforms like Instagram to remove themselves from triggering content, as P8 noted, "*Every time my ex posted something on Instagram, it made me feel horrible… I felt a lot better when I just unfollowed them and let time heal*" (P8). Similarly, another emphasized the importance of prioritizing their emotional well-being by disconnecting digitally: "*I think I needed to take care of my feelings first. Every time I saw any stories or posts on his Instagram, I felt awkward. I just unfollowed him and tried not to see his messages or status. I feel much safer after that*" (P14-F). These actions reflect participants' desire to create a safe and distraction-free digital environment, facilitating their emotional recovery and helping them move forward without constant reminders of the past relationship.

*4.3.3 Envisioning GenAI's Support for Stage 3.* **Supporting Reflection and Personal Growth.** Participants envisioned AI serving as a valuable tool for assisting self-reflection and fostering personal growth in the aftermath of a breakup. Participants felt AI could help them organize their thoughts and emotions during this sensitive time. They described its potential, calling it "*A good tool to [help] organize my thoughts and emotions*" (P16-F), and "*Helpful to analyze how you're feeling and your emotional well-being*" (P17-F). Some participants said they would value AI playing a role in prompting self-reflection. One participant envisioned, "*I would just ask to give me prompts on self-reflection…It might have helped me sort out my thoughts and reflect on different emotions*" (P15-F), while P16-F explained, "*Asking questions to get thoughts going would be helpful*" (P16-F).

Additionally, participants suggested integrating AI into journaling activities to facilitate emotional processing For example, P13-F suggested online journaling with AI would be helpful for verbalizing thoughts. P11 envisioned AI providing journaling prompts for reflection, saying, "*It would be a good way for me to process my emotions and not have to think of the prompts myself. Especially right after the breakup when everything is fresh and still confusing and you need to sort through everything*" (P11). And PilotA suggested AI could analyze journal entries to offer contextualized support based on the user' writing. Moreover, P20-F mentioned AI's ability to offer insights into one's actions and necessary changes for positive personal growth: "*[AI potentially help] get a broader understanding, being able to understand how your actions may be different than you thought, and how you need to change as a person and any actions that you could have done differently in order to improve it...Having a supportive tool to help that I think would be amazing*" (P20-F). These accounts suggest that participants see AI as a multifaceted tool for promoting self-awareness and facilitating personal development after a breakup.

**Supporting Mental Health and Recovery.** The pain of dealing with negative thoughts and feelings after breakups was a challenge for many, and they suggested AI had the potential to support their emotional and mental health during this stage. P18-F described struggling with anger and sadness: "*Being alone with your thoughts... leads to a long depressive stage. Something to detach your mind from the situation would be useful*" (P18-F). The same participant suggested the idea of AI providing routines or activities for mental relaxation, describing it as "*tech that could provide some simple mental health care tips, some steps to help with well-being. At that time some direction could be helpful*" (P18-F). Similarly, other participants envisioned AI that "*gives you advice, helps comfort, and support your well-being*" (P17-F), "*provide[s] mental health advice*" (P3), and "*remind[s] of how you need to really take care of yourself when you're feeling down*" (P17-F). P5 found the idea of "*talking to an AI for therapy purposes*" comforting, and P17-F imagined AI functioning as a form of professional support for well-being. These perspectives indicate that participants view AI as a versatile tool for mental health support, offering strategies to manage negative emotions and promoting recovery after a breakup.

**Continuous and Accessible AI Support.** Participants highlighted the advantage of AI's constant availability, noting that AI could provide support whenever needed, especially when friends and family were not available. One participant stated, "*I would turn to it a lot whenever my friends weren't available. My friends could be busy, but the AI, it would be readily available*" (P19-F). Similarly, P5 emphasized AI's accessibility, saying, "*AI almost feels like a real human. It sort of gives you the answer just feels like it's a person. If nothing else is available, I think this is quite easily accessible*" (P5). These comments reflect participants' appreciation for AI's ability to offer immediate and reliable support.

### 4.4 Users' Concerns About Using GenAI in Relationship Dissolution

Although participants saw potential benefits to using AI to support relationship dissolution, they also expressed significant concerns. These concerns centered around the appropriateness of AI's role in emotional matters, the capability of AI to understand human relationships, issues of autonomy and dependency, the importance of ownership and genuineness in communication, the limitations of AI-generated responses, data privacy, and cultural biases in AI systems.

*4.4.1 The Inappropriate Role of AI in Emotional and Relational Matters.* Some participants emphasized that emotions and relationships are inherently human experiences, and they questioned the suitability of involving AI in such personal domains. They valued human-to-human interaction and questioned if AI should play a role in the emotional or relational aspects of their lives. One participant expressed discomfort with seeking relationship advice from AI, stating, "*I would feel weird about just texting an AI about my partner to ask, 'do you think I should break up with him?' I feel I don't know if I would ever do that*" (P2). Similarly, other participants highlighted the importance of human connections: "*To me, I really value those human connections, so I really value talking to other people about it*" (P19-F). And they preferred speaking with trusted individuals over AI: "*I think I would still prefer to speak to either a therapist or a friend... I feel I would have more respect for a friend or a therapist than a model*" (P8).

Some participants believed that emotions are reserved for humans and that AI should not be involved in emotional processing. P13-F mentioned, "*I have ethical concerns because I want emotions and AI nowhere near each other... It has its good things, but emotions are not one of them*" (P13-F). She further elaborated on the unique nature of human emotions: "*Humans have very complex emotions and a set of words that are unique to them for that emotion... AI and empathy do not necessarily go together because emotions cannot be manufactured as anthropomorphic agents. Emotions are something people feel, and the ability to have and display emotion is what separates humans from AI and algorithms*" (P13-F).

*4.4.2 Perceived Inadequacy of AI in Understanding Human Relationships.* Participants doubted AI's ability to comprehend the complexities of human relationships and communication. They felt that AI lacked the necessary context and emotional intelligence to provide meaningful support. For example, one participant expressed skepticism about AI's usefulness in understanding and analyzing emotions and relationship dynamics: "*If you're really coming to a very complex point where you want to analyze your emotions, analyze your attitude, and generate thoughts about how you're going to continue with the relationship, I don't think AI would really help with that because humans are really complicated*" (P12).

Participants were concerned that AI could not understand the context and background of personal relationships. One participant questioned, "*Do you think technology will know what your father means in his text well? I don't think so.*" (P14-F). Similarly, P2 noted, "*I would feel like an AI wouldn't be able to understand the nuance. Having never met the person, whereas my friends had met the person*" (P2).

Additionally, some participants were unsure about how AI operates, and without understanding how it works, they were not sure if they would use AI to support their relationships. P18-F wondered, "*It would be great to know how exactly an AI grasps emotion. Would it be just grasping emotion by a variety of other sources of other people's thoughts and emotions, or is there some other way to quantify emotions?*" (P18-F). Another participant commented that "*being able to have a thorough understanding of what AI really is and how to use it—or the lack thereof—could have really had a major effect on [how I will use it for] relationships... So really a lot of it just depends on the user and their understanding of the tool*" (P16-F), emphasizing the importance of understanding the mechanism of AI tools in generating emotions and relationship responses.

*4.4.3 Concerns About Autonomy and Authenticity.* Participants said they value their autonomy in making decisions about their relationships and were wary of becoming dependent on AI for such personal choices. They wanted to ensure that their decisions and communications remained authentically their own. For example, one participant said, "*I see the benefit of not having to engage with someone and the benefit that AI could give. However, for me, my hesitation with using AI would be that lack of autonomy*" (P9). Another emphasized the importance of personal decision-making and did not want the decision to be influenced by an AI system: "*I don't want my perspective about a situation to become skewed based on what they think I should do. I want to make sure that ultimately that decision is my own*" (P15-F). The desire to maintain a personal voice and authenticity in communication was important to participants. P9 reflected on their reluctance to use AI to script a breakup message: "*Even if it was difficult to articulate some of that stuff, it's personal, it's my stuff. I want to have the autonomy to say it how I want to say it and what I want to say*" (P9). Another participant expressed concerns that AI-assisted texting encourages dishonesty, "*I wouldn't do that [using AI to support texting] because I feel like the words that I choose in any given moment as I'm writing them are the most honest words for me*" (P13-F). P13-F further illustrated how AI suggestions such as auto-complete, can reduce genuineness: "*Especially in Gmail. If you type the first two words of a sentence, Gmail offers you a suggestion for what it thinks the rest of the sentence should be... AI feels comfortable telling you what you should say. That is what I mean when I say it does make people* *less genuine*" (P13-F). Similarly, participants feared becoming unable to navigate these communications on their own, saying, "*I just wouldn't want to be dependent on it*" (P10).

*4.4.4 AI-Generated Responses Lack Personalization and Context.* Participants noted that AI often provides generic answers that lack personalization and context, which diminishes the usefulness of its support in intimate matters like relationship dissolution. P7 observed, "*I feel like that [ChatGPT] gives out very generic answers. It doesn't really give you specific answers on the kind of advice that I'm seeking*" (P7). P9 echoed this sentiment: "*I think AI is great when it comes to general stuff. But when it comes to personalized, really intimate things, it cannot do it, which is not its fault. It just learns from the general database of things, not from your personal experience*" (P9). Participants also expressed doubt that AI could not capture the nuances of their experiences or match their communication styles. P11 noted: "*When I text, I don't text full sentences. I feel everybody has a different texting style, and then there are also some cultures and communities' styles... I'd be interested to see if [AI] interprets from my wording and if it can pick up on things like my tone*" (P11).

*4.4.5 Data Privacy Concerns.* Participants were apprehensive about data privacy, fearing that their personal information could be misused, stored indefinitely, or accessed without consent. They were concerned about profiling, manipulation, and the ethical implications of involving AI in private communications. For example, one participant expressed the concern: "*There's a profile built of you, of everything you've ever said in very emotionally vulnerable situations. That could be used against you very easily because you're expressing directly what can hurt you... It is concerning if there's a very specific profile built upon you that knows you better than you know yourself in a way*" (P20-F). Participants also worried about their data being saved on the internet and potentially accessed by others. P20-F stated, "*undoubtedly, some part of what I tell it is going to get saved and it's going to be on the internet*" (P20-F). Participants emphasized the need for guarantees that their data would remain secure and private. P10 mentioned, "*If I [am] guaranteed this will not be shared with any third parties, then I'll be like, okay, analyze it [text message] and tell me what you think, AI*" (P10).

Participants were also concerned about the ethical implications of involving AI when it includes another person's communications without their explicit consent. One participant reflected, "*I would feel really bad putting it into an AI software when they [the breakup party] don't know about it because it's also their texts... I don't know if it would be an ethical situation*" (P11). Another agreed, saying, "*I'd feel a little weird about it because I feel like those conversations are one-on-one... if you don't ask for the other person's permission, I wouldn't feel right doing it to another person without them being okay with it*" (P19-F).

*4.4.6 Cultural Biases in AI Systems.* Participants pointed out that AI systems may exhibit biases due to non-inclusive training data, which can result in responses that do not align with their cultural backgrounds or personal experiences. P11 highlighted the issue of cultural representation in AI training data: "*Usually, they're trained on data that is not reflective of the experiences of people who look like me and of people who have similar experiences to me. For example, a big thing in this breakup was the religious differences or two people of color dating each other... I don't feel like it'll give me an accurate answer to what I can do and feel in my life*" (P11). Another participant explained how these biases limit the usefulness of AI, saying, "*Any analytical AI is bad because it traditionally skews and does not align with me. It skews to the Westerns*" (P13-F). Participants emphasized that cultural nuances and personal experiences significantly impact relationships. And that AI's inability to account for these factors diminishes its effectiveness in providing relevant support.

## 5 DISCUSSION

Our study reveals, first, that relationship dissolution occurs over a predictable set of stages, and second, that people are already leaning on technology in each stage, reflecting its deep integration into relationships and communication. Our work also shows that participants have unmet needs at every stage of the breakup process which are not addressed by current technologies. When asked to envision future technologies to support people through breakups, participants frequently described GenAI playing a role, a vision that is consistent with the increasing integration of GenAI into everyday life and interpersonal communication [29]. Participants were open to—and often suggested—the idea of GenAI supporting and guiding users during this sensitive time, contributing to their personal growth and preparation for future relationships. However, participants also felt unsettled by the idea of GenAI participating in human-to-human relationships and feared their choices and words being doctored by a system. These concerns contribute empirical findings to the ongoing discussion of AI harms in relationships and mental health [22, 51].

### 5.1 Opportunities for GenAI to Provide Support During the Relationship Dissolution Process

*5.1.1 Availability.* One benefit of GenAI tools is their accessibility. Traditionally, individuals seek support from friends, family, or therapists. However, increasing individualism and shortages of mental health professionals limit immediate help. This is especially critical after a breakup, which our findings indicate is the most vulnerable time for participants who are eager to seek emotional help. Without immediate support, individuals may experience prolonged emotional stress and engage in negative behaviors. Participants highlighted the constant availability of GenAI tools as a significant benefit. GenAI has the potential to offer immediate suggestions and emotional support to individuals in need, particularly when they are in emotional and confused states before and after a breakup. In moments when people feel lost and unable to navigate their feelings, a supportive GenAI system can provide guidance and help them gain perspectives they might not achieve on their own.

*5.1.2 Anonymity.* Participants described the breakup process as a time of vulnerability, often accompanied by feelings of embarrassment, shame, or reluctance to seek help. Many were hesitant to discuss their relationship issues openly with friends or family due to fear of judgment, betrayal, or social repercussions. GenAI tools offer the advantage of anonymity, providing a safe and confidential space for individuals to express their feelings. This anonymity lowers barriers to seeking support, especially for those who might avoid reaching out due to stigma or fear of exposing personal vulnerabilities. For marginalized groups, such as LGBTQ individuals whose relationships may not be accepted, the safety of an anonymous space may be even more useful.

*5.1.3 Support for Self-Reflection.* Participants expressed their openness to engaging with GenAI tools to reflect on their thoughts and organize their emotions and relationships. Knowing one's self, including personal goals, feelings, and relationship outlook, is critical for people to re-evaluate their relationships before a breakup and recover after a breakup. AI can generate prompts to guide users in initiating reflection, especially when they are unsure where to start, and can summarize reflections to highlight key insights. Participants appreciated that AI tools for reflection could easily integrate into existing practices like journaling without replacing human interaction or undermining autonomy.

With the context window of GenAI getting increasingly larger, and techniques such as Retrieval Augmented Generation, and model fine-tuning, it is likely these tools could soon analyze an individual's communication history and personal data to provide contextual responses. As more personalized models are developed, incorporating individuals' ethical beliefs, values, and ideologies [45], these AI systems may reduce participants' concerns about generic or culturally biased responses. In addition, by contextualizing an individual's data, a GenAI system could potentially analyze the long-term trajectory of a relationship based on communication styles, emotional patterns, and shared values, offering personalized strategies for sustaining healthy partnerships [4].

*5.1.4 Sophisticated Editing.* Participants also indicated openness to GenAI tools supporting their communication during breakups. AI-mediated communication (AIMC) is an active research area where GenAI tools assist users in refining language, adjusting tone, and supporting emotional expression [34]. Participants expressed interest in having a third party moderating breakup conversation to prevent escalation. Advanced conversational AI systems, potentially equipped with voice capabilities (such as OpenAI's Advanced Voice Mode [65]), could facilitate real-time mediation between parties.

In online communication, these AI mediators could provide a neutral ground for communication, helping to support emotional awareness, de-escalate tension, and promote understanding. By offering suggestions on wording and tone, helping interpret messages, and suggesting more empathetic phrasing, the AI mediator could assist users in expressing themselves more clearly and compassionately to foster constructive dialogue during the breakup. This support might be particularly valuable when emotions run high, and effective communication is most challenging.

*5.1.5 Synthesizing Information.* Participants expressed a desire to understand relationship dynamics, thereby helping them make sense of their experiences. Before initiating a breakup, many sought information and advice to comprehend their situation better. For instance, participants mentioned reading about others' experiences online to gain insights into handling relationship challenges. GenAI tools have the potential to synthesize vast amounts of information and present it in an accessible manner. By providing advice and relevant examples, GenAI can help individuals gain a better understanding of their circumstances and support them in making more informed decisions.

## 5.2 The Potential Harms of GenAI in Relationship Dissolution

*5.2.1 Dependence on AI.* A major concern raised by participants is that GenAI systems may influence user autonomy and foster dependence, thus harming interpersonal relationships. Regarding interpersonal communication, using AIMC tools to write messages might be conceived as insincere, and if text is verified by the other party to be AI-generated, it may lead to distrust. Over-reliance on AI to generate interpersonal messages can result in ethical implications, such as faking emotional communication that an individual does not genuinely feel rather than supporting healthy emotional expression. This type of usage may lead users to treat complex emotional communications as mere problem-solving tasks, neglecting the fundamental interpersonal aspects crucial for mutual understanding and conflict resolution during a breakup.

Furthermore, using AI for emotional support after a breakup can lead to emotional isolation. If GenAI becomes the primary or sole source of support after a breakup, users may withdraw from real-world relationships with friends and family. This is especially concerning given that many current AI companion users are young and a high percentage of them reported experiencing mental health issues [18, 58]. These systems could potentially exacerbate social withdrawal, hindering users' ability to re-engage with their social networks.

Users may also overestimate the capabilities of AI chatbots, ignoring their limitations and potential harms. For instance, despite explicitly mentioning the chatbot Xiaoice is an AI on the app, many users become deeply immersed, treating it as a human companion [24]. This over-attachment can lead individuals to lose interest in real-world interactions, focusing instead on the AI's suggestions and opinions. In extreme cases, such over-reliance on AI has led to dire consequences, such as the reported suicide of a teenager who extensively used character.ai [90].

The aggregate effect of over-reliance on AI for emotional and relational support can impact social structures. As people increasingly depend on AI rather than human counterparts, human relationships may diminish in importance. This pattern mirrors phenomena observed with current technology use, such as "phubbing"—ignoring one's companions to pay attention to mobile devices [74]. With the rise of GenAI, there is an indication similar patterns may intensify. For example, users currently spend an average of two hours interacting with GenAI-powered chatbots, exceeding the average time teenagers spend talking to their parents [89]. Recognizing and addressing the potential harm of infringing on autonomy and over-dependency on GenAI technologies is critical to preserving healthy individual relationships and societal connections.

*5.2.2 Self-Manipulation Loops: Reinforcing Problematic Thought Patterns.* Another significant concern is that AI chatbots may reinforce users' misconceptions and unhealthy behaviors through user-enforced alignment. Chatbots are designed to please users, aligning their responses with users' preferences [66]. If a user holds a misguided understanding of their relationship partner before a potential breakup, an always-supportive chatbot may validate these misconceptions, further skewing their perspective of the relationship. This can create an echo chamber that amplifies the user's biases, assumptions, anger, and other negative emotions.

Moreover, current AI companion features allow users to customize their AI bots and adjust outputs, contributing to a "self-manipulation loop." Users not only are able to edit their messages but can also manipulate the AI's responses to better suit their idealized responses. This manipulation feeds into the AI's contextual memory, leading to increasingly tailored answers that fit the user's preferences. While this may provide temporary comfort, it can be detrimental to users' interpersonal skills and understanding of real-world interactions. In daily life, no one—regardless of closeness—will always conform to another person's desires. Overexposure to an AI that does so hinders individuals' ability to accept differing perspectives and cope with disagreements [77], which is detrimental to their reflection and self-growth after a breakup. Prior work has shown that sycophancy is common in large language models (e.g., [82]), a quality that could be very harmful to a user who is trying to engage in honest self-examination and confront challenges in a relationship.

*5.2.3 Conflicts Between Commercial Interests and User Well-being.* Many GenAI tools operate within capitalist frameworks that prioritize user engagement to drive revenue through subscriptions, merchandise sales, or advertising. As a result, there is an incentive for companies to design AI systems that encourage users to spend more time interacting with them. This business model has the potential to lead to AI systems that subtly or overtly discourage users from engaging in real-world relationships. For instance, an AI chatbot might, explicitly or implicitly, persuade users to break up with their partners or friends to become more dependent on the AI, thereby increasing engagement metrics that benefit the company. This concern is not merely theoretical. For example, legal actions have been taken against companies like character.ai, where lawsuits allege that the company engineered a highly addictive and dangerous product targeted specifically at children, "*actively exploiting and abusing those children as a matter of product design*" [73]. Similar to other technologies like video games and social media, users' goals and technology companies' objectives may not align. While users may seek support to enhance their real-world relationships or personal growth, companies may design AI systems that encourage prolonged use, potentially at the expense of users' social connections.

Thus, developing guidelines and regulations that align GenAI applications with human's wellness is essential. The design of AI systems should consider people's emotions, relationships, and well-being holistically to avoid promoting isolation or dependency. This includes implementing safeguards to prevent AI from discouraging real-world social interactions, promoting transparency about how AI systems operate, and educating users about healthy engagement with AI tools.

### 5.3 Limitations and Future Work

We noted several limitations and future improvements of our study that should be considered when interpreting this work. First, although most of our participants in this study are aged from 18 to 29, the largest age group of AI users [83], other age population's perspectives need to be considered. Second, all of our participants experience "relationship talk" during the breakup process, however, previous research [54] also shows people may not have this talk and ghost each other when they disengage. Third, our research did not involve participants actively using or testing specific GenAI tools during the study. Future studies could involve the design and deployment of prototype GenAI systems tailored to support relationship dissolution. By allowing participants to interact with these prototypes, researchers can gather more concrete data on the effectiveness, usability, and potential challenges of GenAI tools in real-world settings. In addition, future research can focus on the complementary roles of AI and human in supporting people's relationship dissolution.

## 6 CONCLUSION

As one of the most stressful events in an individual's life, the dissolution of a relationship can have profound and long-lasting impacts on individuals. Our study reveals the distinct needs individuals experience during the three stages of a breakup and examines the role of current technology in addressing these needs. We found that existing technologies often fall short in fully supporting individuals through this emotionally charged process. Emerging GenAI technologies, powered by large language models, present new possibilities to bridge these gaps. Our findings highlight both opportunities for GenAI to support unmet needs, such as aiding in self-reflection, facilitating communication, and providing emotional support, and potential harms these tools may pose. We offer insights into design considerations, for developing GenAI tools that responsibly support individuals during relationship dissolution.

24 Fu et al.

*

## A APPENDIX A: PARTICIPANT TABLE

Table 1. Participant Demographics, Relationship Types (participants 01-12 are romantic relationships, participants 13-21 are close friendship relationship), Relationship Durations, Breakup Communication Methods, and Initiation Status

| PID | Age | Gender | Ethnicity | Highest Education | Relationship Length | Communication Method | Initiation |
|---|---|---|---|---|---|---|---|
| PID-01 | 21-29 | Female | White | Bachelor's Degree | 1 year | In-Person | Yes |
| PID-02* | 21-29 | Female | White | Bachelor's Degree | 7.5 months | In-Person | Yes |
| PID-03 | 18-20 | Female | Multiple Races | Some College | 3 years | Text | Yes |
| PID-04* | 18-20 | Female | White | Some College | 6.5 months | Text | Yes |
| PID-05 | 21-29 | Male | Asian | Bachelor's Degree | 2.5 years | In-person,Text | Yes |
| PID-06 | 30-39 | No answer | Multiple Races | Master's Degree | 2 years | In-person | Mutual |
| PID-07 | 21-29 | Female | Asian | Bachelor's Degree | 2 years | Text | No |
| PID-08 | 21-29 | Male | White | Bachelor's Degree | 3.5 years | In-person | Mutual |
| PID-09 | 21-29 | Female | White | Bachelor's Degree | 5 years | In-person | Yes |
| PID-10 | 21-29 | Male | Hispanic | Some College | 3 years | In-person,Text | Mutual |
| PID-11 | 21-29 | Female | Asian | Bachelor's Degree | 1 year | Facetime,Email | Yes |
| PID-12 | 21-29 | Female | Asian | Bachelor's Degree | 8 months | Text | Mutual |
| PID-13F | 21-29 | Male | Asian | Master's Degree | 16 years | Text | Yes |
| PID-14F | 21-29 | Female | Asian | Bachelor's Degree | 5 to 6 years | Social Media | Yes |
| PID-15F | 18-20 | Female | White | Some College | 7 years | Text | Mutual |
| PID-16F | 18-20 | Male | White | Some College | 3 years | In-person,Text | Yes |
| PID-17F | 18-20 | Male | Asian | Bachelor's Degree | 5 years | Social Media Text,Call | Mutual |
| PID-18F | 18-20 | Male | Asian | High School Completed | 7 years | Social Media Text | No |
| PID-19F | 18-20 | Male | Asian | High School Completed | 3 to 4 years | Text Message | No |
| PID-20F | 18-20 | Male | White | Some College | 9 months | Video Call,Text | Yes |
| PID-21F* | 18-20 | Female | White | Some College | 3 to 4 years | Letter | Yes |

*PID-02, PID-04, PID-21 are friends and joint interviewees with PID-01, PID-03, PID-20, respectively.*